\documentstyle[multicol,epsf,pre,aps]{revtex}

\begin{document}
\date{\today}
\draft

\title{Universal Power Law in the Noise
from a Crumpled Elastic Sheet.}
\author{Eric M. Kramer\cite{byline} and Alexander E. Lobkovsky}
\address
{The James Franck Institute and the Department of Physics\\
The University of Chicago\\
Chicago, Illinois 60637}
\maketitle

\begin{abstract}

Using high-resolution digital recordings, we study
the crackling sound emitted from crumpled sheets
of mylar as they are strained.
These sheets possess many of the qualitative
features of traditional
disordered systems including frustration and discrete memory.
The sound can be resolved into
discrete clicks, emitted during rapid changes in the rough
conformation of the sheet. Observed click energies range over
six orders of magnitude.
The measured energy autocorrelation function for the sound
is consistent with a stretched exponential
$C(t) \sim \exp (-c \, t^{\beta})$ with $\beta \approx .35$.
The probability distribution of click energies has a power law
regime $p(E)\sim E^{-\alpha}$ where $\alpha \approx 1$.
We find the same power law for a
variety of sheet sizes and materials,
suggesting that this $p(E)$ is universal.

\end{abstract}

\pacs{68.60.Bs, 75.10.Nr}

\begin{multicols}{2}
Physicists and laymen alike are familiar with the loud
crackling noise produced when a crumpled cellophane candy
wrapper is unwrapped. This noise is a generic product
of thin, crumpled elastic sheets and is qualitatively
different from the sounds that can be produced with a
flat sheet. Despite its ubiquity and accessibility,
to our knowledge there has been no experimental
investigation of this sound.
In this paper we present
an elementary discussion of the generation of the sound and its
dependence on the properties of the crumpled sheet.
We develop a method to generate the sound
which yields robust and reproducible results and
we make a quantitative study of the sound properties
using high-resolution digital recordings.

The key distinction between flat and crumpled
sheets is the quenched curvature disorder produced
during the crumpling process. At high compressions
some regions of the material become sharply
creased \cite{ridge}, and the material at the
fold flows plastically to relax the strain.
The creases thus initiate line-like defects where the sheet
prefers a nonzero curvature. If the sheet is subsequently
uncrumpled it exhibits a rough texture due to the presence
of these defects, which appear as a network of ridges in the sheet.
For this reason we expect that a systematic investigation
of the sound produced by these sheets will be relevant
to a variety of issues in the theory of crumpled membranes.
Thin crumpled sheets of paper and aluminum foil
have been studied informally
by several authors \cite{crmpl}, and the theory of
elastic sheets with quenched disorder is an active field of
research \cite{glassy}. Our research is also
closely related to the work of Houle and Sethna,
who study the sound produced when a flat
sheet is crumpled \cite{HS}.

Our first task was to make elastic sheets with quenched disorder
in a uniform and reproducible way. We start with a flat sheet
of mylar \cite{mylar}, and successively crumple and uncrumple
the sheet by hand. The result after one crushing
is unsatisfactory for our purposes, since there
are long-range
correlations in the ridge network and we have doubts
about the reproducibility of the result \cite{crmpl}.
However, after thirty crumplings there is a dense
network of overlapping ridges
and an additional crumpling produces
no obvious change in the appearance or the sound of the sheet.
The final state is homogeneously rough down to a lower
length scale $\xi$, which is limited by our ability to crush the sheet.
If we estimate $\xi$ for our samples using the size of the
largest flat facets, we find $\xi \approx 100 h$
where $h$ is the thickness of the sheet.

The disordered sheets produced in this way have
an unusual property. Manipulating a sheet
by hand, we can find many inequivalent stable configurations.
Fig. 1 illustrates three configurations for the {\it same} sheet.
These configurations can hold their shape for at least months.
The reason for this behavior is the quenched curvature disorder.
Fig. 2 represents the crossing in the sheet of two ridges
from two distinct crumpling events.
Each ridge has a preferred dihedral angle $D \ne\pi$, but only one
can be achieved. The energy of the pair is thus explicitly
frustrated \cite{Toul}.
A sheet of linear size $L$ will have ${\cal O}(L^{2}/\xi^{2})$
crossings. In this way the elastic energy of the sheet resembles the
free energy of a spin glass. This is in accord with the suggestions
of several investigators that membranes
with quenched curvature disorder should exhibit glassy
behavior at low temperature \cite{glassy}.
The multitude of stable configurations is
a clear demonstration that the elastic energy of the sheet
has many distinct minima. Typical energy barriers between minima
are much greater than $k_{B}T$ for a macroscopic
sheet at room temperature.

Our recordings were made using an AKG C34
microphone calibrated to have a linear response
for frequencies from 20 Hz to 20 kHz. The analog signal
from the microphone
was converted to a 16-bit digital signal at a 48 kHz sampling rate
by a Yamaha ProMix 01 mixer and recorded to digital audio tape.
The recordings were made in a music studio that
was neither soundproof nor

\begin{figure}
\centerline{\epsfxsize=8cm \epsfbox{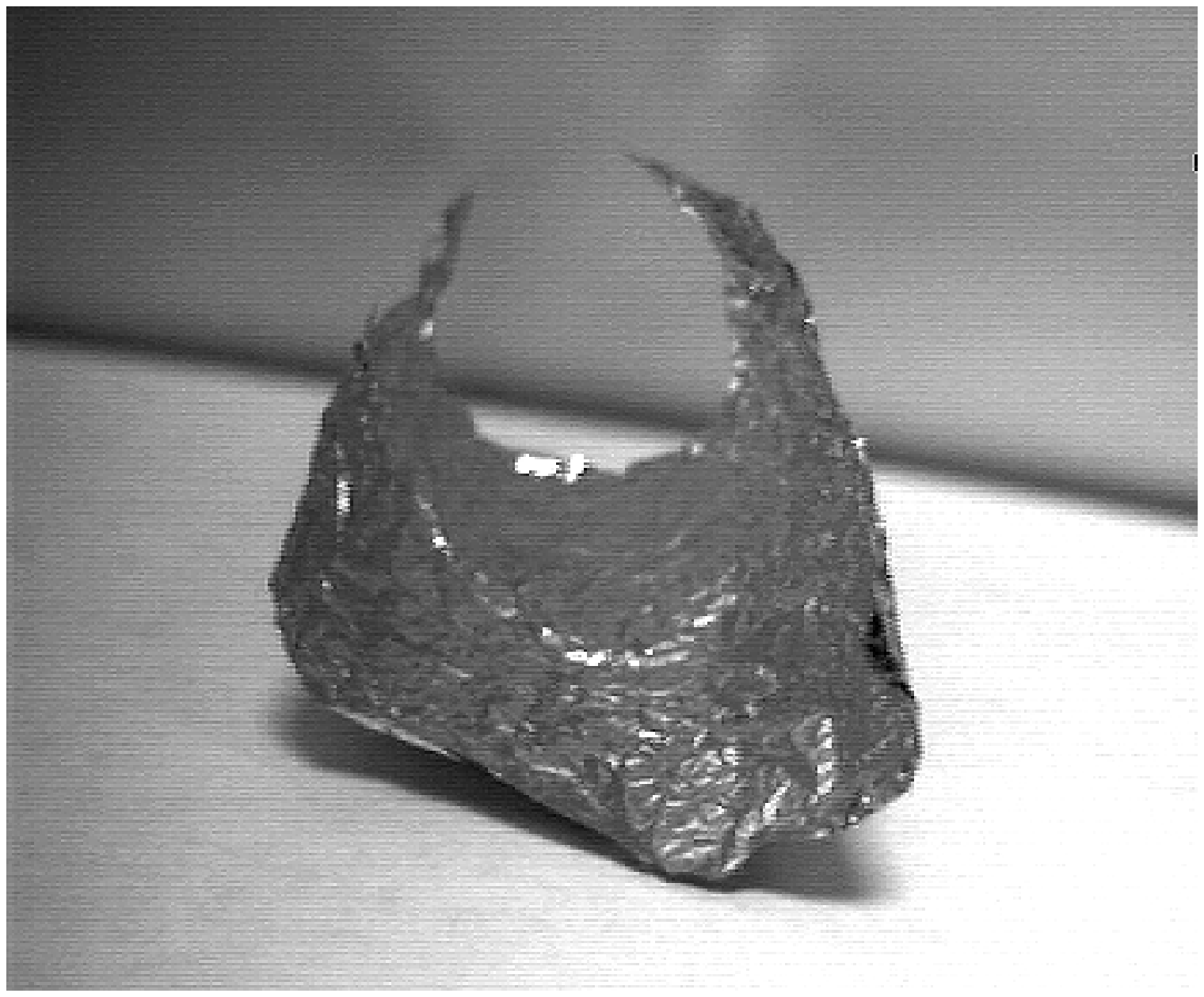}}
\centerline{\epsfxsize=8cm \epsfbox{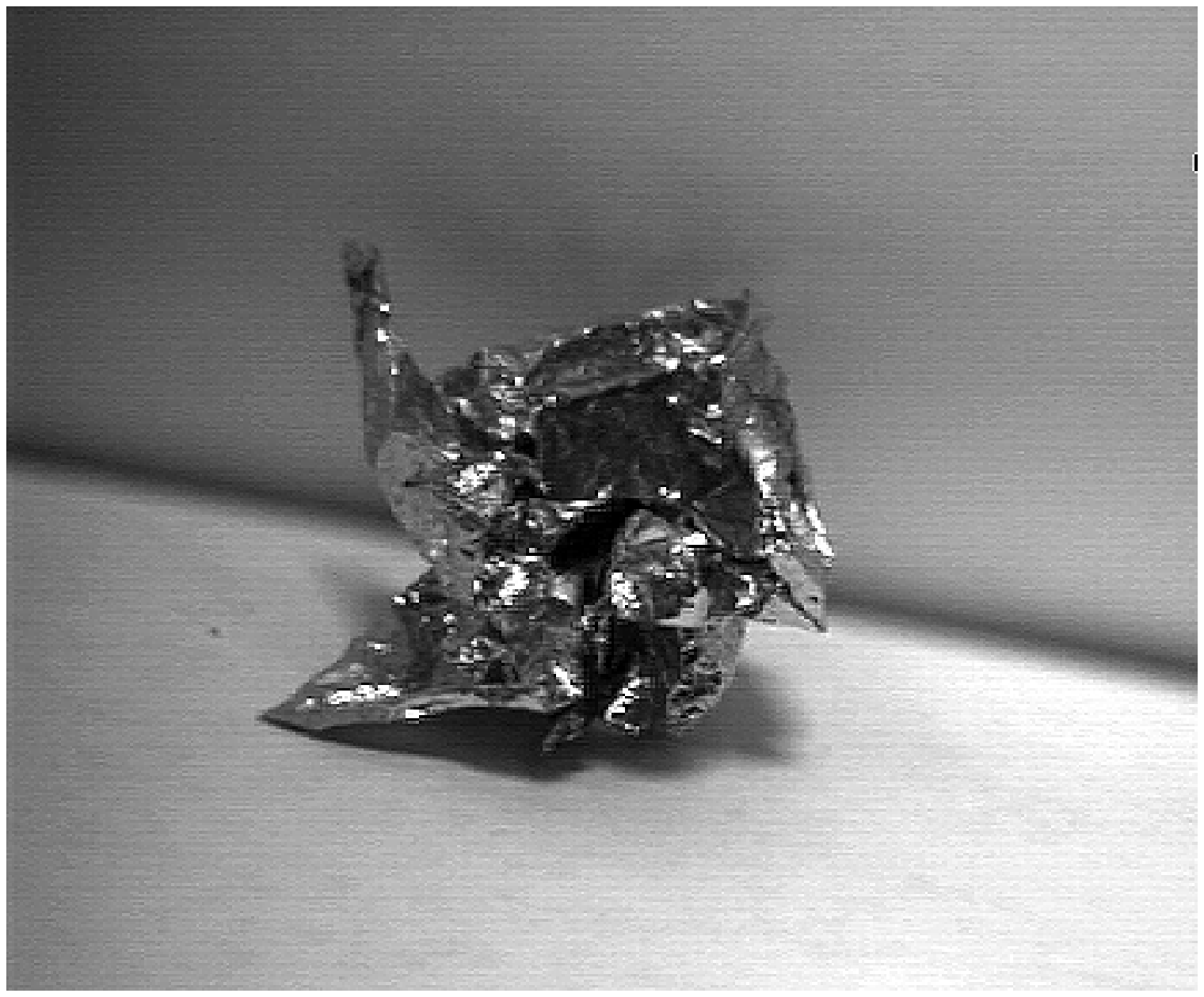}}
\centerline{\epsfxsize=8cm \epsfbox{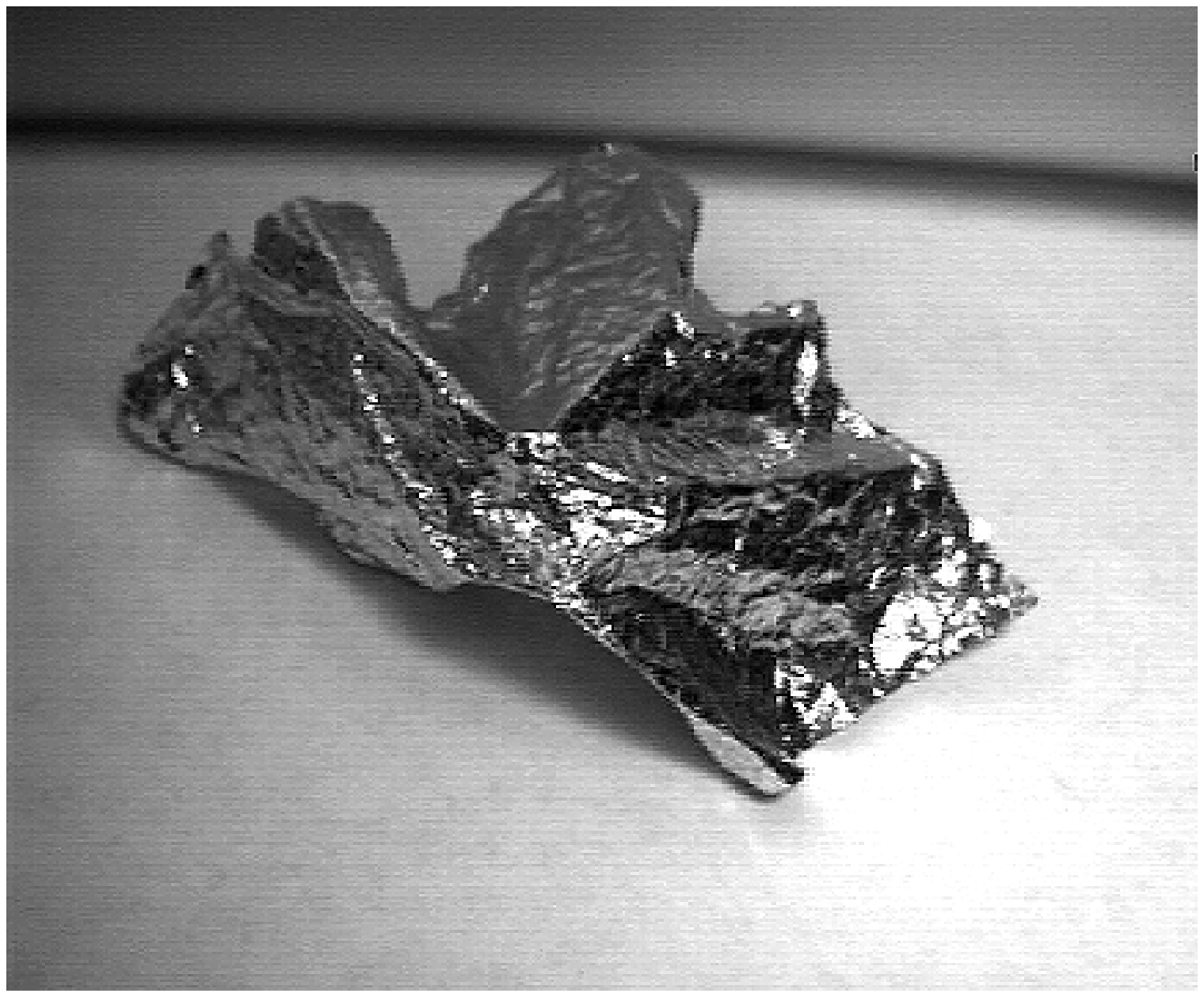}}
{FIG. 1. Three stable conformations of a mylar
sheet with quenched curvature disorder. The dimensions
of the sheet are 8 cm x 9 cm x 0.5 mil.}
\end{figure}

\noindent anechoic.
Since ambient room noise
was only significant at frequencies below 200 Hz, we used
the mixer as a high-pass filter with an
attenuation of 15 dB below 100 Hz and no attenuation
above 400 Hz. Most of the energy in our

\begin{figure}
\centerline{\epsfxsize=8cm \epsfbox{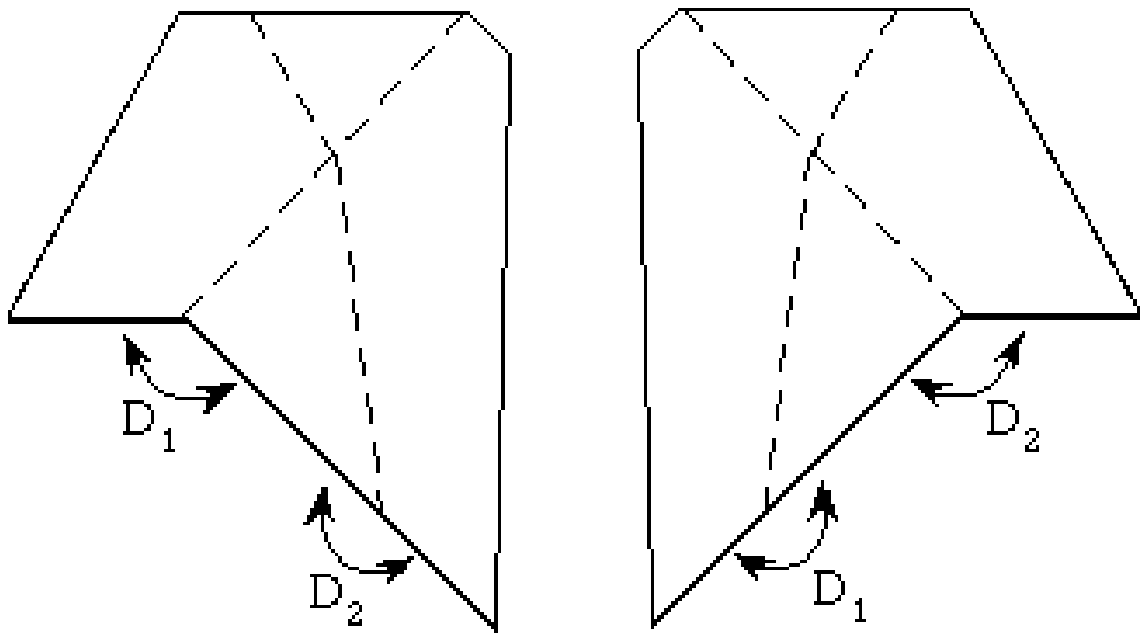}}
{FIG. 2. Schematic illustration showing that one of two crossed
ridges must have $D=\pi$.}
\end{figure}

\begin{figure}
\centerline{\epsfxsize=8cm \epsfbox{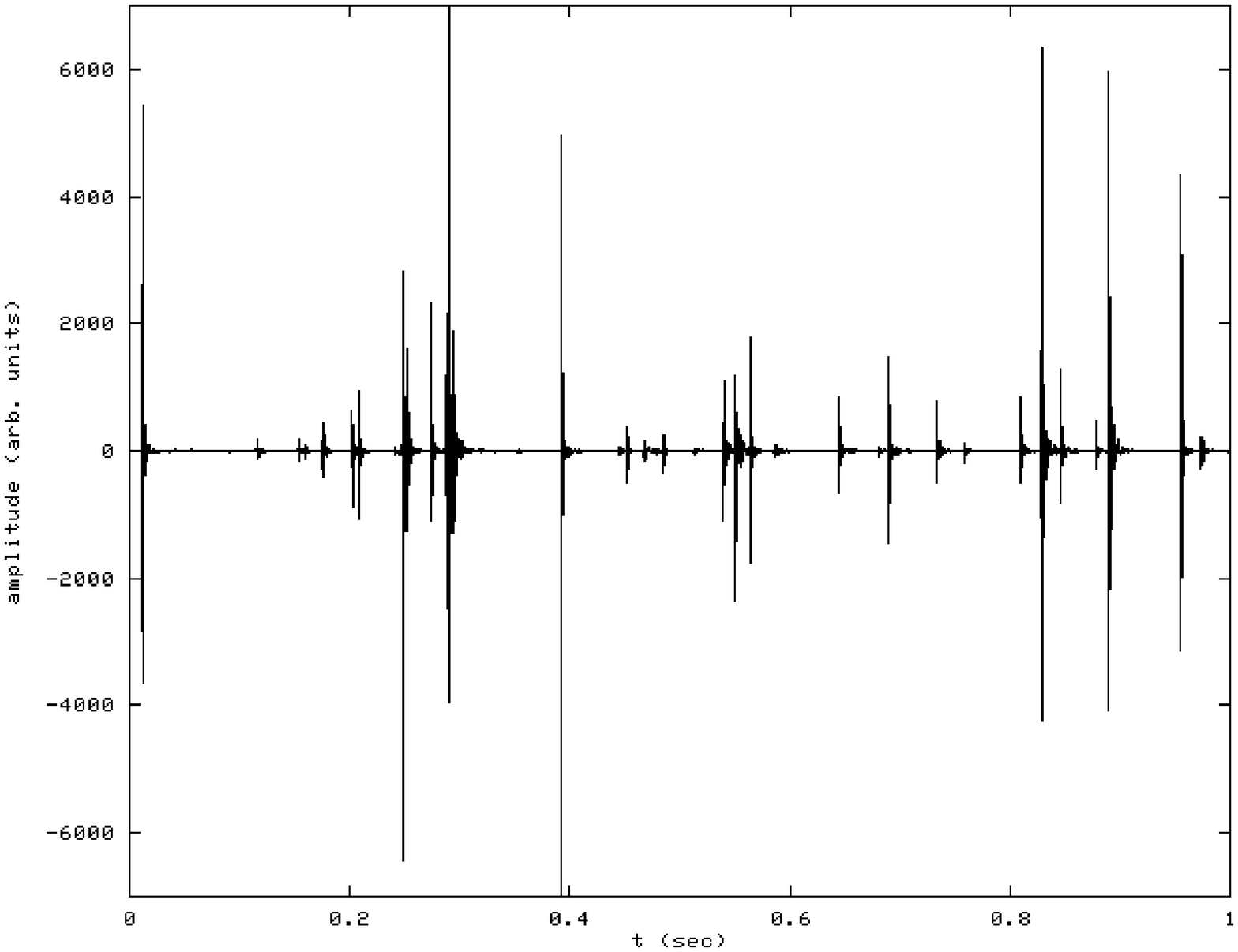}}
{FIG. 3. Amplitude vs. time for 1 second
of sound data from a sheet 3 mil thick.}
\end{figure}

\begin{figure}
\centerline{\epsfxsize=8cm \epsfbox{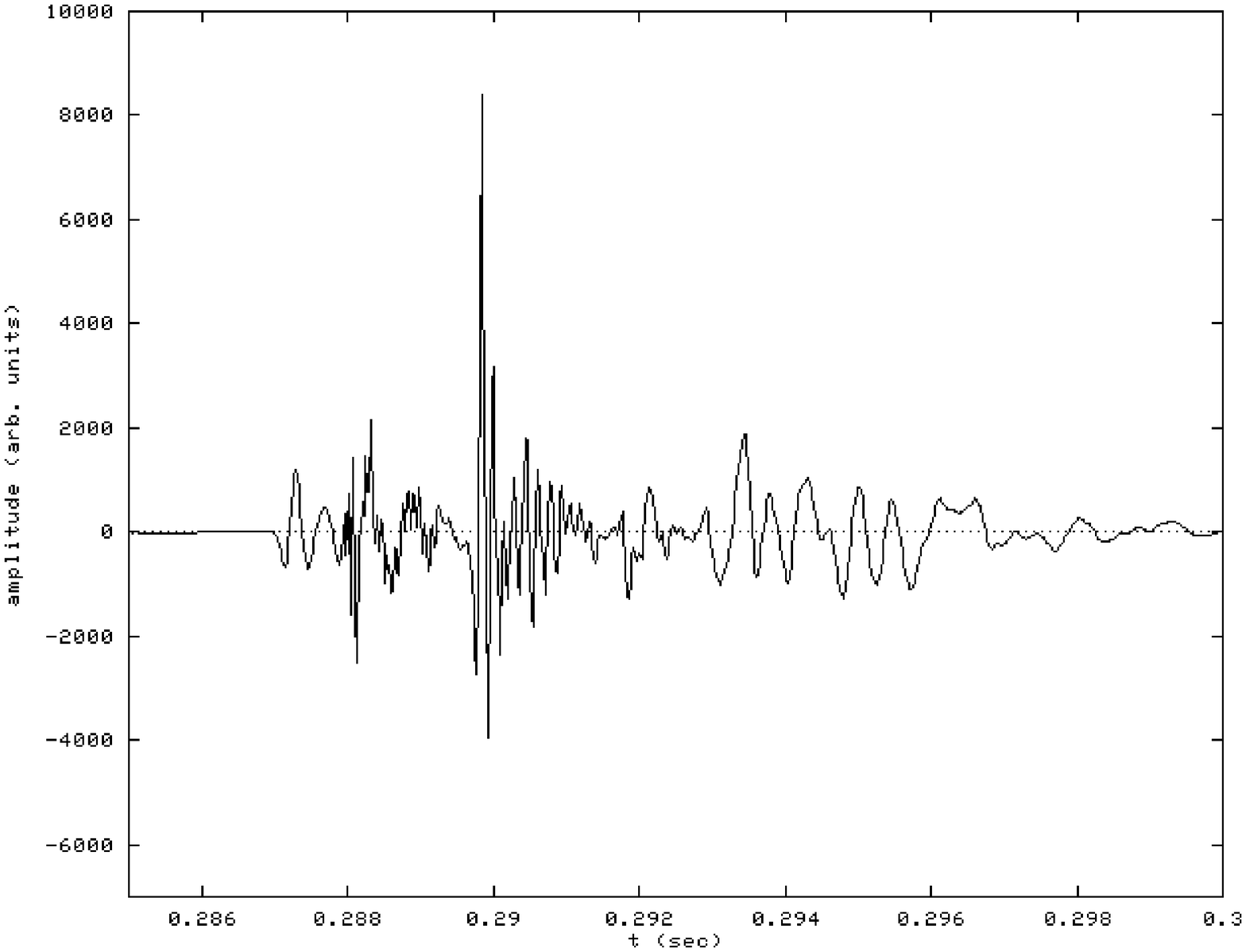}}
{FIG. 4. The large click near $t = 0.3$ sec in Fig. 3.}
\end{figure}

\noindent signal
was above 1 kHz, so this was an adequate solution.
To overcome the problem of echoes from the studio walls
we constructed a soundbox by covering the inside
of a cubical cardboard box 20 in on a side with
3 in of dense foam rubber packing material.
One wall of the box was left open so that we could
hold the crumpled sheets inside.
We verified that the soundbox decreased the measured
amplitude of echoes to negligible levels.

The results presented here were recorded from a set
of five square mylar sheets prepared in a rough state
as described above. The flat sheets were initially
10 cm on a side and 0.5, 1.0,
3.0, 5.0 and 10.0 mil thick respectively.
We produced the sound by holding a sheet by its edges
in an approximately flat configuration
and slowly twisting and shearing it in a cyclical fashion.
Typical shearing rates were order 0.5 cm/sec.
It was important to keep the sheet approximately flat
to avoid two complications. (1) If the sheet rubbed against
itself, it produced low amplitude friction noise
that diminished the sensitivity of our recordings.
(2) If the sheet crumpled significantly, we would need
to account for the attenuation of sound from the inner layers.
The sheets were held about 20 cm from the
microphone and each run was 10 seconds long.
We made ten runs in all, with each investigator
taking a turn straining each sheet.
To get the best use of the full energy
scale, the gain of the mixer was changed for each run.
We also made a 3 sec run for all
five sheets with the gain fixed, so that we could
compare the results from different sheets.

Fig. 3 shows one second of sound data recorded from a
3 mil sheet. The amplitude $i(t)$ is proportional to
pressure, so the energy in the signal is $E(t) \sim i^{2}(t)$.
The sound is clearly resolved into many discrete
{\it clicks}. The integrated energy of the clicks
ranges over at least six orders of magnitude. Fig. 4 shows
a single large click from the same run. This click appears
to be a superposition of several distinct vibrational
modes.

The source of the sound is evident if one examines
the sheet while straining it. Each click is emitted
during a sudden change in the rough conformation.
The visible change is often localized
to a region of size $\xi$, and can
be ascribed to the buckling of a single facet bounded
by ridges. The largest clicks originate from
the buckling of regions much larger than
the roughness length scale. They can probably be interpreted
as the superposition of many smaller events. This may be the
origin of the internal structure seen in Fig. 4.
Some clicks demonstrate a discrete memory effect.
By alternately stretching and compressing the sheet,
a single buckling event can be induced
and reversed, emitting a click in both the forward
and reverse portions of the cycle. Under favorable
circumstances, this cycle can be repeated dozens of times.

These observations support a simple picture for the mechanism
of sound generation in the disordered sheet.
According to the spin-glass analogy, the disordered
sheet has many nearly degenerate energy minima corresponding to
distinct rough conformations \cite{glassy}.
The work done to strain the sheet
is stored as elastic energy.
When the total work done is enough to overcome the
potential barrier to a neighboring minimum, the
sheet buckles and the stored energy is released as
heat, vibrations in the material, and the sound of a click.
This picture is reminiscent of the Barkausen effect,
the magnetization jumps during a field reversal in the
random-field Ising model, and acoustic emissions during
microfracturing processes \cite{Bark,RFIM,micro}. These
are all examples of externally driven systems with quenched disorder,
and it is natural that the sound from a disordered elastic
sheet should be included in this category.

The five runs made at fixed gain let us make
a comparison of the sound from different
thicknesses of mylar. Because we had limited
control of the straining process
and because the cross-sheet comparisons rely on absolute
measures of the energy, the results
in this paragraph should be
regarded as qualitative only. A detailed discussion
of the data will be avoided.
At a constant strain rate the energy in the largest
clicks increases with the thickness $h$
like $h^{2}$ while the total power in the sound
stays roughly constant.
This implies that the number of clicks per second decreases
like $h^{-2}$.
The upward trend in the energy of individual clicks
is probably due to the energy
scale $Yh^{3}$ associated with the material,
where $Y$ is the Young's modulus.
The downward trend in the number of clicks
can be accounted for
by the postulate that the number of clicks emitted at
fixed strain rate is proportional
to the density of frustrated ridges in the sheet
$N_{r} \sim (L/\xi)^{2} \sim (L/h)^{2}$.

Fig. 5 shows the
 normalized energy autocorrelation function
\begin{eqnarray}
C(t) = \frac{\langle E(t^{\prime})E(t^{\prime}+t)\rangle
-\langle E \rangle ^{2}}{\langle E^{2} \rangle -\langle E \rangle ^{2}}
\end{eqnarray}
for one run each of $h$ = 0.5, 1.0, 3.0, 5.0, and 10.0 mil.
Note the cusp at the origin.
A log-log plot of the data (inset)
demonstrates that the autocorrelation function is
not adequately described as an algebraic decay.
A log-linear plot (not shown) demonstrates that
it is also not an exponential decay.
Fig. 6 shows that
all five curves are well described by a stretched
exponential $C(t) \sim \exp (-(t/t_{0})^{\beta})$.
A fit by eye gives $\beta = .3-.4$ and
$t_{0} = (3-5) \, 10^{-2}$ msec.
Stretched exponential
correlation functions are commonly found for systems
with glassy relaxation dynamics \cite{micro,stretch}.
Our autocorrelation functions all achieve a minimum in the vicinity
of $t \approx 10$ msec, and thereafter exhibit fluctuations
of magnitude .01. We have confirmed that these late-time
($t \gg t_{0}$) fluctuations are due to the poor
statistics associated with the largest and
least frequent clicks in the sample.

The most interesting characteristic of the sound
is $p(E)$, the probability per unit time
to encounter a click of energy $E$.
One way to measure this would be to write an algorithm
which identifies individual clicks. Unfortunately, we find that
the computer identification of smaller clicks is an
ambiguous task.
Instead, we work with the running sum
$E_{\Delta}(t)\equiv \sum^{t+\Delta}_{t^{\prime}=t} E(t^{\prime})$
and define $p(E;\Delta)dE$ as the probability that
$E \le E_\Delta < E+dE$.
The estimate for the click energy distribution
$p(E) \approx p(E;\Delta)/\Delta$
is valid when (i) $\Delta\gg t_{0}$, (ii)
$\int^{E}_{0} dE^{\prime} E^{\prime} \, p(E^{\prime};\Delta) \ll E$,
and (iii) $\int^{\infty}_{E} dE^{\prime}\, p(E^{\prime};\Delta) \ll 1$.
The first condition ensures
that a click isn't counted twice. The second condition

\begin{figure}
\centerline{\epsfxsize=8cm \epsfbox{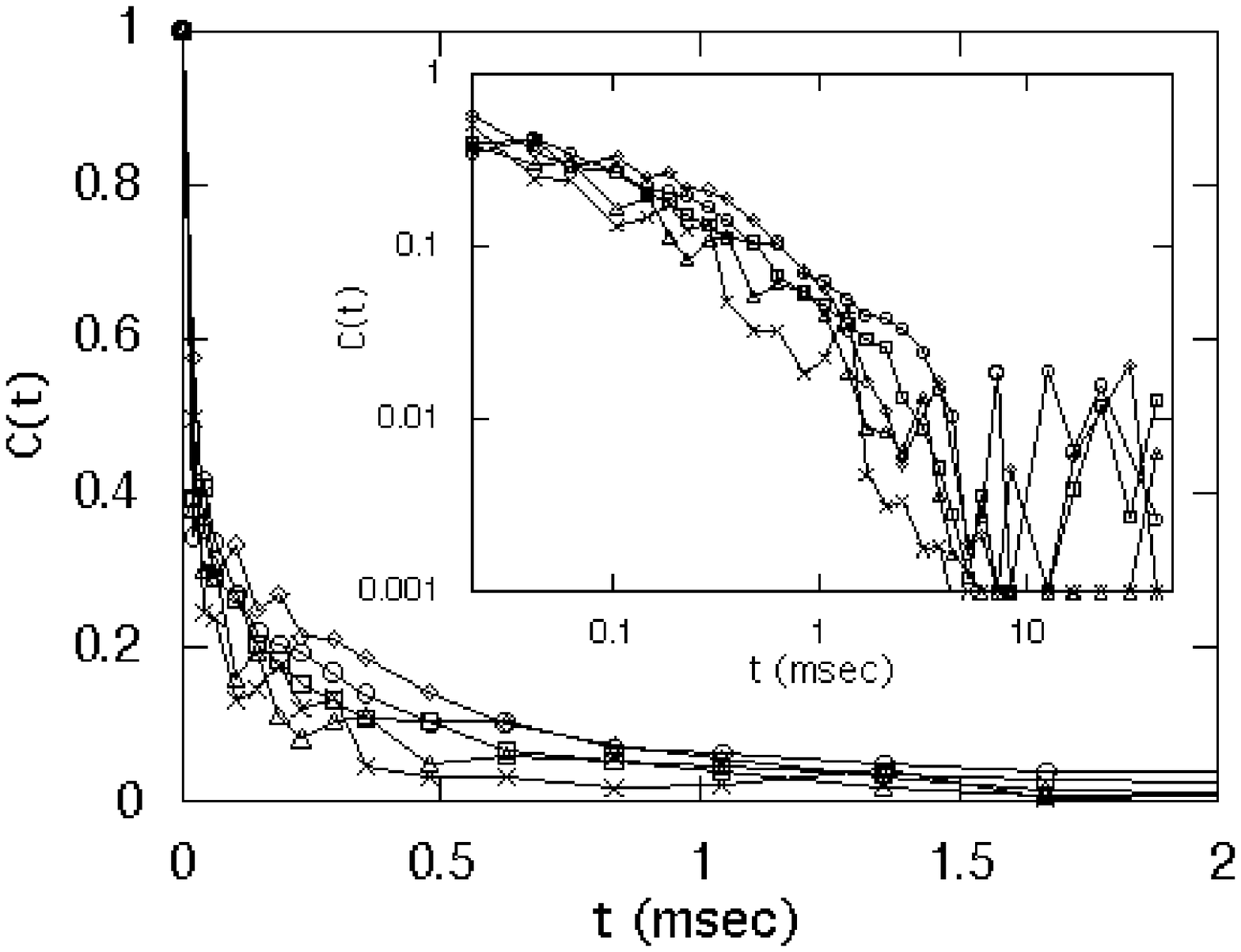}}
{FIG. 5. The energy autocorrelation function for
$h$ = 0.5 ($\bigcirc$), 1.0 ($\Box$), 3.0 ($\Diamond$),
5.0 ($\triangle$), and 10.0 mil ($\times$).
Inset: log-log plot of the same data.}
\end{figure}

\begin{figure}
\vspace{-.15in}
\centerline{\epsfxsize=8cm \epsfbox{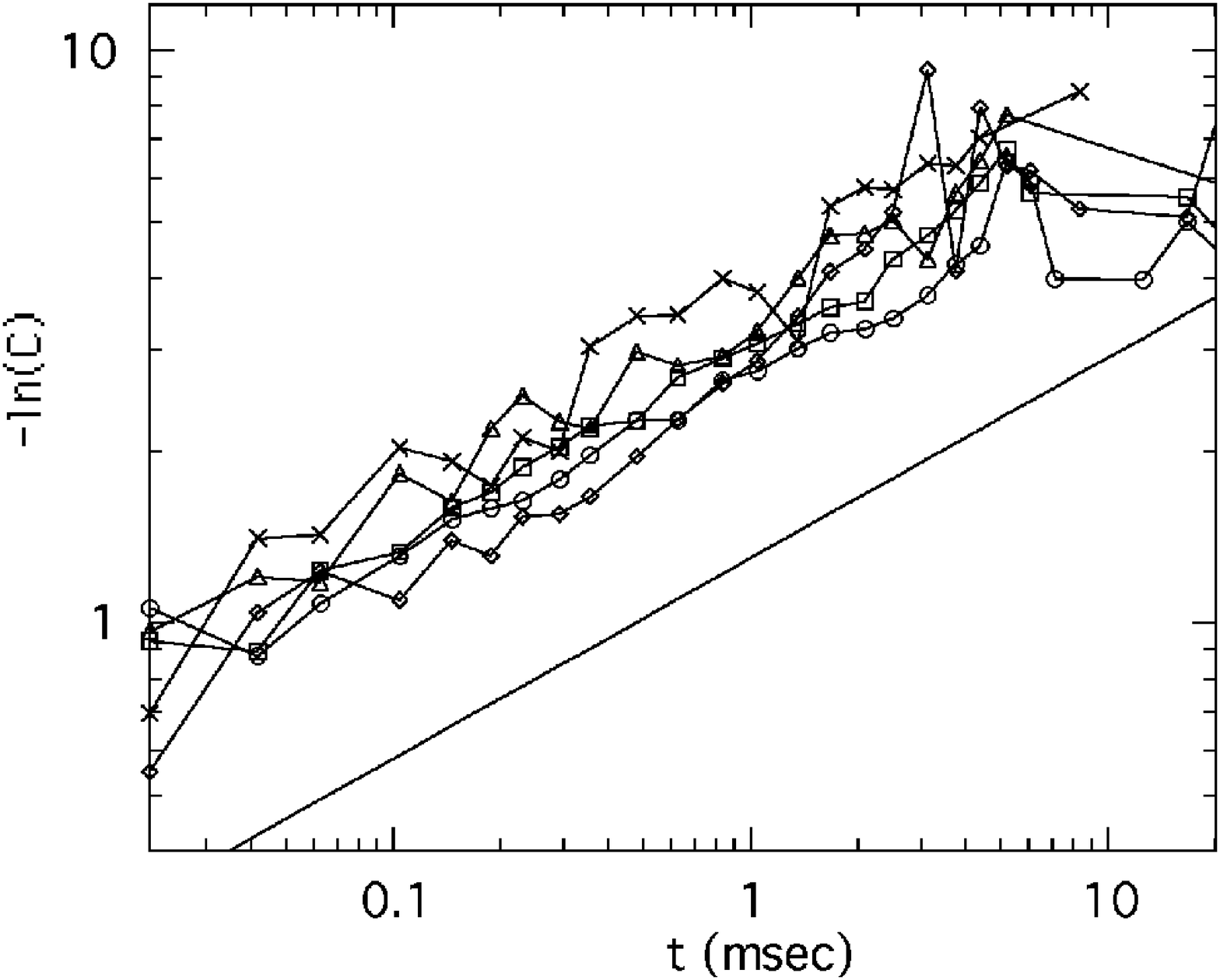}}
{FIG. 6. Plot of -ln(C) (symbols same as in Fig. 5).
The straight line is a stretched
exponential with $\beta=.35$.}
\end{figure}

\begin{figure}
\vspace{-.15in}
\centerline{\epsfxsize=8cm \epsfbox{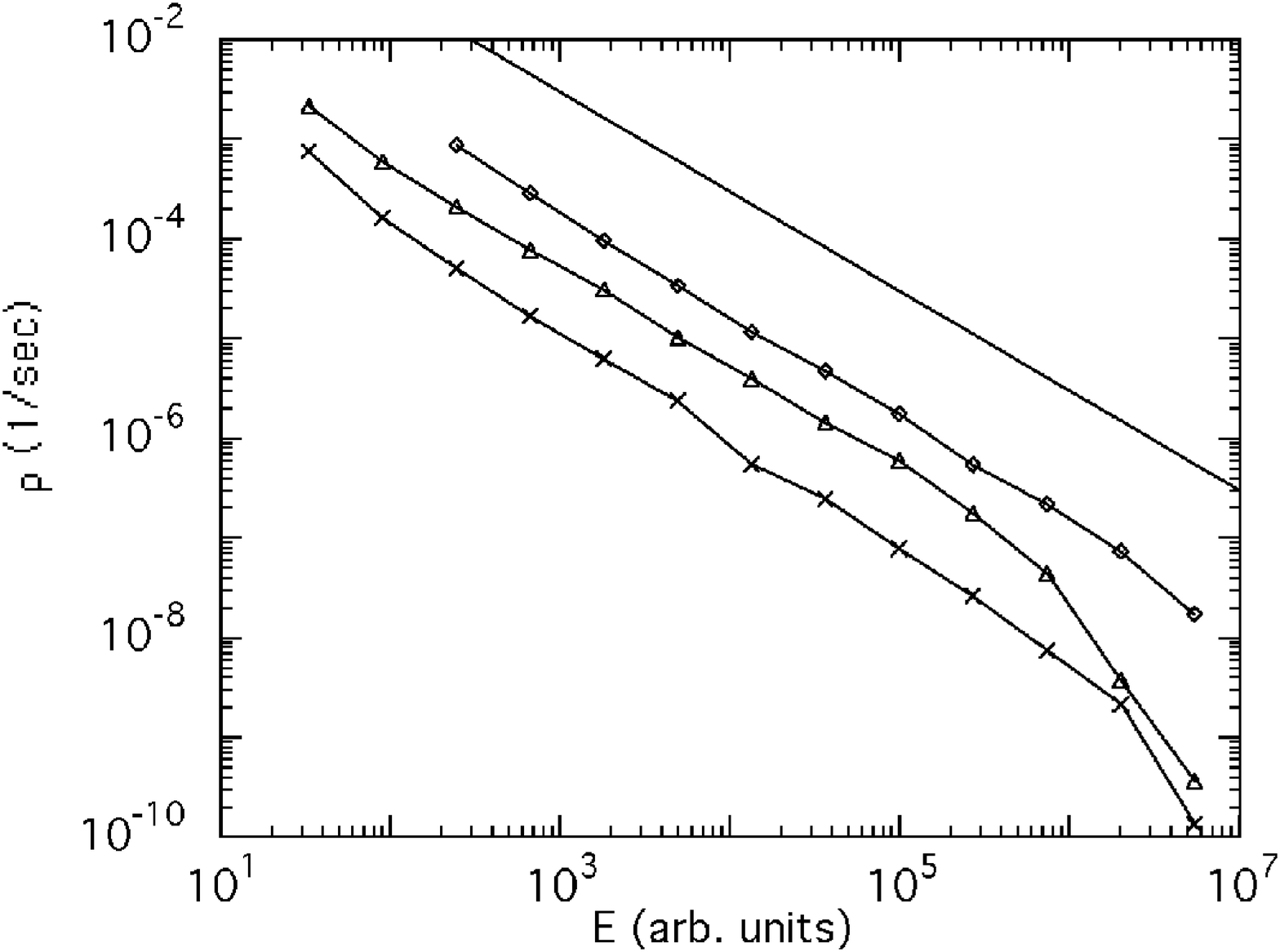}}
{Fig. 7. Plot of the click energy distribution function
for $h$ = 3.0 ($\Diamond$),
5.0 ($\triangle$), and 10.0 mil ($\times$). The straight
line is a power law with $\alpha=1.0$.}
\end{figure}

\noindent ensures
that the value of $E_{\Delta}$ is dominated by the {\it largest}
click in $\Delta$. The third condition ensures that
large clicks aren't so common that they hide the majority
of smaller ones.

Fig. 7 shows our results for $p(E)$ for one run
each of thicknesses $h$ = 3.0, 5.0, and 10.0 mil.
We have chosen $\Delta=10$ msec to satisfy condition (i).
Due to low-amplitude noise in the signal,
conditions (ii) and (iii) fail when
$p \gtrsim 0.001$.
We see that the click energy distribution exhibits a
power law $p(E) \sim E^{-\alpha}$ over four decades
in the energy. A least squares fit gives $\alpha =1.1 - 1.0$.
For the $h$ = 0.5 and 1.0 mil sheets the clicks
were too numerous. Conditions (ii) and (iii) weren't satisfied
and $p$ couldn't be determined.
Although our best results are presented here, we have
also examined the sound from a variety of crumpled sheets
made from materials other than mylar and with a
variety of different sizes.
In all cases where $p$ can be determined
we find a power law regime
in the distribution of click energies
with an exponent ranging from 1.2 to 1.0. We suggest
that this result is universal.

One motivation for examining the click energy distribution
$p(E)$ is to probe the length scales associated with the
crumpled sheet.
Our initial expectation was that there would be
a single energy scale, determined by the roughness length $\xi$.
The result $p(E) \sim 1/E$ implies there is no
typical energy scale in the crumpled sheet.
This may be a reflection of many relevant dynamical length scales
in the system. If this is the case, then the upper cutoff
of the click energy distribution is determined by the finite
size of the sheet and the lower cutoff is determined
by $h$ or $\xi$. However, there is a second possible source of
variability in click energy. The distribution of elastic
stresses in a rough configuration is likely to be
extremely inhomogeneous.
The variation in click energies
over six orders of magnitude is probably due to both effects.

The most obvious criticism of our experiment is that
we strain the sheets by hand, so the reproducibility
of our results is in question.
Nevertheless, we can borrow
an argument from the theory of isotropic turbulence
and note that the strains on the sheet are applied
at a length scale $l \sim 1$ cm,
which is much larger than the length scale $\xi$
at which the strains are relaxed.
Similarly, the {\it longest}
time scale measured for the clicks is 10 msec. This
is much faster than the macroscopically slow changes
in the boundary conditions.
We therefore expect
that the detailed nature of the driving force is
irrelevant for our results.
We are supported in this by the robust
nature of the click energy distribution $p \sim 1/E$.
Experiments under more controlled conditions should resolve
this issue.

In conclusion, we have demonstrated that
elastic sheets with quenched curvature
disorder have a variety of glassy characteristics
including many distinct energy minima,
discrete memory, and stretched exponential relaxation.
The sound produced by these sheets is a direct
consequence of the complex energy landscape
associated with glassy systems.
This paper raises a number of interesting
questions. (1) The most pressing experimental issue
is the refinement of the experimental procedure
and the verification of these results under
more controlled circumstances. (2) One would like
to better understand the mechanism of click generation,
to assess the extent to which the energy in the
sound of a click represents the elastic
energy released during a buckling event.
(3) Lastly, can the theories already developed for
elastic sheets with quenched disorder account for
the stretched exponential autocorrelation function
and the power law distribution of click energies?

\acknowledgments
The authors would like to thank Tom Witten, Sid Nagel,
and David Grier for helpful discussions. We would also like to
thank John Hungerford-Shirley for his support in the
recording studio. This work was supported in part
by the NSF through Grants No. DMR-9205827 and DMR-9400379.

\end{multicols}
\end{document}